%Paper: hep-th/9508089
%From: M.Kneipp@swansea.ac.uk
%Date: Fri, 18 Aug 95 11:33:39 +0100

\magnification=1200 \hsize=6.2 truein \vsize=8.9 truein
\def\ni{\noindent}
\hfill SWAT/94-95/81
\medskip
\ni{\bf EXACT ELECTROMAGNETIC DUALITY}
\bigskip
\ni{\bf David I Olive}\bigskip
\ni Dept of Physics, University of Wales Swansea
\medskip\ni{\it Invited Talk given at the Trieste Conference on Recent
 Developments in Statistical Mechanics and Quantum Field Theory, April 1995}
\bigskip
\noindent{\bf Introduction}\bigskip
Electromagnetic duality is a very old idea,
 possibly predating Maxwell's equations. Although the route that has recently
led to
 a precise and convincing formulation has been long, it has turned out
to be of quite surprising interest. This is because it has synthesised many
hitherto
independent lines of thought, and so intriguingly interelated disparate ideas
arising in the quest for a unified theory of particle physics valid in the
 natural space-time with three space and one time dimension.
Despite the progress, final proof is lacking and likely to
 require further breakthroughs in fundamental mathematics.

Although nature does not seem to display  exact electromagnetic duality,
realistic theories could well be judiciously broken versions of the exact
theory
in
which sufficient structure survives to explain such
 long-standing puzzles as quark confinement in the way advocated by
Seiberg and Witten [1]. Spectacular support for their arguments comes from
 applications in pure mathematics where new insight has been gained into
 the classification of four-manifolds [2], transcending
 the celebrated work of Donaldson [3].

Here I shall review the developments leading up to the formulation
of exact electromagnetic duality, taking the view that an
 understanding of this must precede that of the symmetry breaking.
\bigskip
\noindent{\bf The Original Idea}
\bigskip
\def\epb{{\underline E}+i{\underline B}}
The apparent similarity between the electric and magnetic fields
${\underline E}$ and ${\underline B}$ was confirmed and made
more precise by Maxwell's discovery of his equations.
 In vacuo, they can be written concisely as just two equations [4]:
$$\nabla.(\epb)=0,\eqno{(1a)}$$
$$\nabla\wedge(\epb)=i{\partial\,\,\,\over\partial t}(\epb)\eqno(1b)$$
at the expense of introducing a complex vector field $\epb$.
These equations display several symmetries whose physical
 importance became clear subsequently. They display Poincar\'e
 (rather than Galilean) symmetry, and, beyond that, conformal symmetry
(with respect to space-time transformations preserving angles
and not just lengths). Unlike the Poincar\'e symmetry,
the conformal symmetry is specific to four space-time dimensions.
Even more sensitive to the precise space-time metric is the electromagnetic
duality rotation symmetry of Maxwell's  equations
$$\epb\rightarrow e^{i\phi}(\epb)\eqno(2)$$
since only in $3+1$ dimensions do the electric and magnetic fields both
constitute vectors so that the complex linear combination $\epb$
appearing in (1) and (2) can be formed.
It is the extension of the fascinating symmetry (2) of (1) that
 is the main theme of what follows.

Notice that we can form two real, quadratic expressions
invariant with respect to (2) [4]:
$${1\over 2}|\epb|^2={1\over 2}(E^2+B^2),$$
$${1\over 2i}(\epb)^*\wedge(\epb)=\underline E\wedge\underline B,$$
respectively the energy and momentum densities of the electromagnetic field.

On the other hand, ${1\over2}(\epb)^2$ is complex
 with real and imaginary parts given by
$${1\over2}(E^2-B^2)+i\underline E.\underline B.$$
As the real part is the Lagrangian density, this shows
 that it forms a doublet under (2) when combined with
 $\underline E.\underline B$ which is a total derivative.
Thus the Maxwell action forms a doublet with a
\lq\lq topological quantity" which is proportional to
 the instanton number in non-abelian theories.

We would like to generalise the electromagnetic duality rotation
symmetry (2) to include matter. We could also consider
 generalisations to non-abelian gauge theories of the
 type which seem to unify the fundamental interactions.
In either case we meet the same difficulty that
 the gauge potentials enter the equations of motion
 and that we do not know how to extend the transformation
(2)  to include them. Eventually we shall find a way
 of combining the two generalisations, thereby extending
the symmetry.

There is another, familiar, difficulty with the equations of
 motion in non-abelian gauge theories; that they are conformally invariant
(in $3+1$ dimensions). As a consequence, the gauge particles which
are the quanta of the gauge potentials, should be massless. This is fine
for the photon, but not for any other gauge particles.
This problem seems to be rather general and deep:  unified
 theories chosen according to geometric principles tend to exhibit unwelcome
conformal symmetry. This occurs in string theory too,
at least as a world sheet symmetry. As a consequence, there is a general
problem of understanding the origin of mass through
a geometrical mechanism for breaking conformal symmetry.
 We know of only two possible solutions, at first sight different,
but related in what follows.

The first is the idea that mass arises from the vacuum
 spontaneously breaking some of the gauge symmetry via
 a \lq\lq Higgs" scalar field [5,6,7]. The second is a principle
due to Zamolodchikov [8]  that we now discuss.
\bigskip
\ni{\bf Zamolodchikov's Principle and Solitons}
\bigskip
The second insight into the origin of mass comes from another area of
physics. Yet, as we shall see, it seems connected to the first, in
four dimensions at least. The conformal symmetry on the world sheet
needed for the internal consistency of string theory hinders the
emergence of a physically realistic mass spectrum in an otherwise
unified theory. However, when string theory is abstracted to the study
of \lq\lq conformal field theory", and applied to the study of second
order phase transitions of two dimensional materials, it is seen that
the simple application of heat breaks the conformal symmetry
controlling the behaviour at the critical temperature. In specific
models it was learnt by Onsager, Baxter and their followers [9] that it is
possible to supply heat while maintaining integrability (or
solvability). Zamolodchikov [8] has elevated this observation to a
principle which has rationalised the theory of solitons and more. In
two dimensions, the local conservation laws characteristic of
conformal symmetry (or augmented versions such as W-symmetry) are
chiral. This means that the densities are either left moving or right
moving (at the speed of light) and so can be added, multiplied or
differentiated. If conformal symmetry is judiciously broken,  a
certain (infinite) subset of the chiral densities remain conserved,
although no longer chirally so. Their conserved charges, that is their space
integrals, generate an infinite dimensional extension of the Poincar\'e
algebra in which the charges carry integer spins. The charges with
spin plus or minus $1$ are the conventional light cone components of
momentum. The  sinh-Gordon equation illustrates this nicely.
It can be written
$${\partial^2\phi\over\partial t^2}-{\partial^2\phi\over\partial x^2}+
{\mu^2\over2\beta}\left(e^{\beta\phi}-e^{-\beta\phi}\right)=0.\eqno(3)$$
The last term, proportional to $e^{-\beta\phi}$, can be multiplied by
a variable coefficient $\eta$ so that $\eta=1$ yields (3), while
$\eta=0$ yields the Liouville equation. Liouville exploited the
conformal symmetry of his equation in order to solve it completely, long
ago.

It is interesting to investigate the behaviour as $\eta$ varies from
$1$ to $0$. As long as $\eta>0$, a simple redefinition of the field
$\phi$ by a displacement restores the sinh-Gordon form (3) but with
$\mu$ replaced by $\mu\eta^{1/4}$. As $\mu\hbar$ is the mass of the
particle which is the quantum excitation of $\phi$, we see that it is
singular as $\eta$ approaches zero with critical exponent $1/4$. So we
see how mass arises from the breaking of conformal symmetry.
This short discussion was classical but it extends to the
quantum regime as envisaged by Zamolodchikov [8].

The sine-Gordon equation is obtained from the sinh-Gordon equation (3)
by replacing $\beta$ by $i\beta$. It then
exhibits the symmetry
$$\phi\rightarrow \phi+{2\pi\over\beta}\eqno(4)$$
and, consequently, possesses an infinite number of vacuum solutions
$\phi_n={2\pi n\over\beta}$, $n\in Z\!\!\!Z$, all with the same minimum energy,
zero. The particle of mass $\mu\hbar$ describes fluctuations about any
of these vacua. But there also exist classical solutions which interpolate
two successive vacua and which are stable with respect to fluctuations.
These solutions can be motionless, describing a new particle, the
soliton, at rest, or can be boosted to any velocity less than that of
light. The jump in $n$, equal to $\pm1$, can be regarded as a
topological quantum number, indicating either a soliton or an
antisoliton. What is particularly remarkable is that one can consider
a solution with an arbitrary number of solitons and/or antisolitons,
initially well separated, but approaching each other, then
colliding and finally emerging with  velocities unchanged
and energy profiles generally unscathed except for time advances relative to
uninterrupted
trajectories [10]. Thus the solitons persist in their structure despite
their collisions and can legitimately be regarded as providing
classical models of a particle with a finite mass and a structure of
finite extent. This phenomenon is a very special feature of
sine-Gordon theory that can be ascribed to the infinite number of
conservation laws mentioned previously, themselves relics of conformal
symmetry.

This sort of integrable field theory has two \lq\lq sorts" of
particle, the quanta of the fluctuation of the field $\phi$ (obtained by
second quantisation) and the solitons which are classical solutions.
Skyrme [11] was
the first to ask whether these two \lq\lq sorts" of particle are
intrinsically different and found that they were not. His explanation  was
that, in the full quantum theory, it is possible to construct a new
quantum field whose fluctuations are the solitons. The new field operator is
obtained by
an exponential expression in the original field $\phi$
%% FOLLOWING LINE CANNOT BE BROKEN BEFORE 80 CHAR
$$\psi_{\pm}(x)=e^{i\beta(\phi\pm\int_{-\infty}^x\,dx'{\partial\phi\over\partial
 t})}\eqno(5)$$
with two spin components (and a normal ordering understood).
 Coleman and Mandelstam [12,13] later confirmed that
$\psi$ satisfied the equations of motion of the massive Thirring model.

The construction (5) is an example of the vertex operator construction
later to be so important in string theory and in the representation
theory of infinite dimensional algebras (resembling quantum field
theories).

There are  multicomponent generalisations of the
sine-Gordon equations called the affine Toda theories, likewise
 illustrating Zamolodchikov's principle in a nice way, and revealing
the role of algebraic structures such as affine Kac Moody algebras
[14].
 Again there are
particles created by each field component, now possessing interesting mass and
coupling patterns (related to group theory). Remarkably, there are an
equal number of soliton species and these display very similar
properties [15].

\bigskip
\ni{\bf Magnetic Charge and its Quantisation}
\bigskip
Let us now return to the question of extending the electromagnetic
duality rotation symmetry (2) to matter carrying electric and magnetic
charges. Suppose first that matter can be regarded as being composed
of classical point particles carrying typical electric and magnetic
charges $q$ and $g$, say. Then it is easy to include the source charges on
the right hand side of Maxwell's equations (1) and to supplement (1)
by the equations of motion for the individual particles subject to a
generalised Lorentz force. This system maintains the symmetry (2) if,
in addition,
$$q+ig\rightarrow e^{i\phi}(q+ig).\eqno(6)$$
The price to be paid for achieving this is the inclusion of unobserved
magnetic charge. We must therefore suppose that the failure to observe
magnetic charge is either due to an associated very large mass or some
other reason.

Turning from the classical to the quantum theory, we immediately find
a difficulty, namely that the electromagnetic couplings of the matter
wave functions require the introduction of gauge potentials, a procedure which
is
not straightforward in the presence of magnetic charge.

Nevertheless, in 1931, Dirac overcame this difficulty and showed that
the introduction of magnetic charge could be consistent with the
quantum theory, provided its allowed values were constrained [16]. His
result was that a magnetic charge $g_1$, carrying no electric
charge, could occur in the presence of an electric charge $q_2$, like
the electron carrying no magnetic charge, provided
$$q_2g_1=2\pi n\hbar\qquad\qquad n=0,\pm1, \pm2,\dots\eqno(7)$$
As he pointed out, this condition had a stunning consequence: provided
$g_1$ exists somewhere in the universe, even though unobserved, then any
electric charge {\it must} occur in integer multiples of the unit
${2\pi\hbar\over g_1}$, by (7). This quantisation of electric charge is
indeed a feature of nature and this explanation is actually the best yet
found. Although apparent alternative explanations, evading the
necessity for magnetic charge, have appeared, they turn out to be
unexpectedly equivalent to the above argument, as we shall see.

There is a problem with the Dirac condition (7), namely that it does
not respect the symmetry (6). In fact (7) is not quite right because,
although Dirac's argument is impeccable, there is an implicit
assumption hidden within the situation considered. It took a surprisingly long
time to rectify this and hence
restore the symmetry (2) and (6), as we see later.

Given that this difficulty is overcome, we can seek a consistent
quantum field theory with both electric and magnetic charges. Then,
presumably, the particles carrying magnetic charge would have a
structure determined by the theory, and hence a mass dependent on the
charges carried. Just as the Maxwell energy density respected the
symmetry (2), we would expect this mass formula to respect (6) so that
$$M(q,g)=M(|q+ig|)=M(\sqrt{q^2+g^2}).\eqno(8)$$
We now proceed to find such a theory.\bigskip\ni{\bf Magnetic
Monopoles (and dyons) as Solitons}
\bigskip
We can now draw together several clues in the ideas already discussed.
One concerns the quantisation of electric charge: since the electric
charge operator, $Q$, is the generator of the $U(1)$ gauge group of
Maxwell theory, its quantisation could be explained by supposing that
it is actually a generator of a larger, simple gauge group (that could
unify it with other interactions). If the larger group were $SU(2)$,
for example, $Q$ would then be a generator of an $SU(2)$ Lie algebra, that is,
an internal angular momentum algebra. Consequently its eigenvalues would be
quantised, thereby providing an alternative explanation of electric
charge quantisation which apparently  evades the need for magnetic
charge.

However, we still have to furnish a mechanism selecting the direction
of $Q$ amongst the three $SU(2)$ directions. This can be achieved by a
scalar field with three components
$\left(\phi_1(x),\phi_2(x),\phi_3(x)\right)$, like the $SU(2)$ gauge
fields. This scalar field has to have the unusual feature of not
vanishing in the vacuum, so that it can select the $Q$ direction
there. It is therefore a \lq\lq Higgs" field providing the mechanism
whereby the vacuum spontaneously breaks the $SU(2)$ gauge symmetry
down to the $U(1)$ subgroup [5,6,7]. As well as this, it also breaks conformal
 symmetry, introducing mass for two of the gauge particles, leaving
the photon massless.

There is a simple formula for the resultant masses of the gauge
particles
$$M(q,0)=a|q|,\eqno(9)$$
where $q$ is the eigenvalue of $Q$ specifying the electric charge of a
specific mass eigenstate. $a$ constitutes a new fundamental parameter
specifying
the magnitude
of the vacuum expectation value of the scalar Higgs field. Actually
the mass formula (9) is much more general. Instead of
the gauge group being $SU(2)$, it could have been any simple Lie
group, $G$, say, and (9) holds as long as the Higgs field lies in the
adjoint representation of $G$, like the gauge fields.

In the vacuum, the gauge group $G$ is spontaneously broken to a
subgroup
$$U(1)_Q\times K/Z,\eqno(10)$$
where $Q$ still generates the invariant $U(1)$ subgroup commuting with $K$.
 The denominator $Z$ indicates a  finite cyclic group
in which the $U(1)_Q$ intersects $K$, and will not be important for
what we have to say.

But this setup, a spontaneously broken gauge theory with a Higgs in
the adjoint representation, is very much the analogue in four
dimensions of the sine-Gordon theory in two dimensions. Instead of the
symmetry relating the degenerate vacua being discrete, (4), it is now
continuous, being the gauge symmetry, $G$, and it is again possible to trap
nontrivial
topologically stable field configurations of finite energy.
Indeed in 1974, 't Hooft and Polyakov [17,18] found a classical soliton
solution emitting a $U(1)$ magnetic flux with strength $4\pi\hbar/q$
in the $SU(2)$ theory with heavy gauge particles carrying
electric charges $\pm q$. Thus there is a soliton
which is a magnetic monopole whose charge indeed satisfies the Dirac
condition (7). Thus the desired novelty of this explanation of electric charge
quantisation is illusory as it reduces to Dirac's original argument [16].
For a more detailed review of the material in this section see [19].

However what we have done is inadvertently achieve our other aim, that of
constructing a
theory in which the magnetic monopoles have structure and a definite
mass which can be calculated by feeding the field configuration into
the energy density and integrating over space. The result is the
following inequality, known as the \lq\lq Bogomolny" bound [20],

$$M(0,g)\geq a|g|.\eqno(11)$$
The similarity to the Higgs formula (9) prompts the question as to
whether the inequality in (11) can be saturated to give equality. This
is possible in the \lq\lq Prasad-Sommerfield" limit in which the
self interactions of the Higgs field vanish [21]. Then the lower bound in
(11) is achieved if the fields satisfy certain first order
differential equations, known as the \lq\lq Bogomolny equations" [19]
$${\cal E}_i=0,\qquad\qquad{\cal D}_0\phi=0,\qquad\qquad {\cal B}_i=\pm{\cal
D}_i\phi,\eqno(12)$$
where ${\cal E}_i$ and ${\cal B}_i$ denote the nonabelian electric and
magnetic fields. Solutions to (12) have zero space momentum and
therefore describe a magnetic monopole at rest, with mass $a|g|$ (if
$|g|$ has its minimum least positive value).

The sine-Gordon solitons satisfy similar first order differential
equations that imply that the mass can also be expressed as a surface
term, but there is an important difference. This is that the Bogomolny
equations (12) (unlike the first order sine-Gordon equations) can also
be solved for higher values of the topological charge, here magnetic
charge. When  the magnetic charge is $m$ times its least positive
value, the space of solutions to (12), called the moduli space, form a
manifold of $4m$ dimensions. $3m$ of these dimensions can be
interpreted as referring to the space coordinates of $m$ individual
magnetic monopoles of like charge.

This means that $m$ like magnetic monopoles can exist in arbitrary
configurations of static equilibrium (unlike $m$ sine-Gordon solitons,
which must move). So, as they have no inclination to move
relatively,
 like magnetic monopoles at rest must fail to exert
forces on each other [22]. (This is reminiscent of the multi-instanton
solutions to self-dual gauge theories: indeed the Bogomolny equations
(12) can be interpreted as self-dual equations in four Euclidean
dimensions).

The remaining $m$ coordinates, one for each monopole, have a more
subtle, but nevertheless, important interpretation: they correspond to
degrees of freedom conjugate to the electric charge of each monopole. Because
of
this, it is possible for each magnetic monopole soliton to carry an
electric charge, $q$, say [23]. In this case, they are called \lq\lq
dyons", following the terminology introduced by Schwinger [24]. Then the
mass of an individual dyon is given by, [25],
$$M(q,g)=a|q+ig|=a\sqrt{q^2+g^2}.\eqno(13)$$
The first remarkable fact about this formula is that it is universal.
It applies equally to the dyon solitons of the theory and to the gauge
particles, as it includes the Higgs formula (9). In fact it applies to
all the particles of the theory created by the fundamental quantum
fields, as it also includes the photon and Higgs particles which are
both chargeless and massless. Thus, whatever $G$, (13) unifies the
Higgs and Bogomolny formulae and is therefore democratic in the sense
that it does not discriminate as to whether the particle considered
arises as a classical soliton or as a quantised field fluctuation [26].

Secondly the mass formula (13) does indeed respect the
electromagnetic duality rotation symmetry (6) as it has the structure (8).

\bigskip
\ni{\bf Electromagnetic Duality Conjectures}
\bigskip
We have seen that spontaneously broken gauge theories with adjoint
Higgs (in the Prasad-Sommerfield limit) have remarkable properties, at
least according to the naive arguments just outlined. Magnetically
neutral particles occur as quantum excitations of the fields present
in the action, whereas magnetically charged particles occur as
solitons, that is, solutions to the classical equations of motion. Yet,
despite this difference in description, all particles
enjoy a universal mass formula (13).

Skyrme showed that, in two dimensions, the soliton of sine-Gordon
theory could be considered as being created by a new quantum field
obeying the equations of motion of the massive Thirring model [11,12,13]. Thus
the same quantum field theory can be described by two distinct
actions, related by the vertex operator transformation (5). It is
natural to ask if something similar can happen in four dimensions,
with the theory under consideration. There, the solitons carry
magnetic charge with an associated Coulomb magnetic field. This
suggests that the hypothetical quantum field operator, creating the
monopole solitons, should couple to a \lq\lq magnetic" gauge group
with strength inversely related to the original \lq\lq electric" gauge
coupling because of Dirac's quantisation condition,
$$q_0\rightarrow g_0=\pm{4\pi\hbar \over q_0},\eqno(14)$$
or possibly half this.

Thinking along these lines, two more specific conjectures were
proposed in 1977. First, considering a more general theory, with a
simple exact gauge symmetry group $H$, (i.e. not of the form (10)),
Goddard, Nuyts and Olive  established a non-abelian version of the
Dirac quantisation condition (7) and used it to propose the conjecture
that the magnetic, or dual group $H^v$ could be constructed in two
steps as follows [27].

(i) The Lie algebra of $H^v$ is specified by saying that its roots
$\alpha^v$ are the coroots of $H$ :-
$$\alpha\rightarrow\alpha^v={2\alpha\over \alpha^2}.\eqno(15a)$$
(ii) The global structure of the group $H^v$ is specified by
constructing its centre $Z(H^v)$ from that of $H$, $Z(H)$ :-
$$Z(H)\rightarrow Z(H^v)={Z(\tilde H)\over Z(H)},\eqno(15b)$$
where $\tilde H$ is the universal covering group of $H$, that is, the
unique simply connected Lie group with the same Lie algebra as $H$.

This conjecture remains open. Notice the similarity between (15) and (14).
In order to make progress, Montonen and Olive sought a  more specific
proposal in a simpler context, and considered spontaneously broken gauge
theories of the
type discussed above, but with the gauge group henceforth definitely
chosen to be $SU(2)$ [26]. This is broken to $U(1)$ by
 a triplet Higgs field so that the mass formula (13) holds good.

The possible quantum states of the theory carry values of $q$ and $g$
which form an integer lattice when plotted in the complex $q+ig$ plane (with
Cartesian coordinates (q,g)). Ignoring possible dyons, the single
particle states correspond to five points of this lattice. The photon
and the Higgs particle correspond to the origin (0,0), the heavy gauge
particles $W^{\pm}$ to the points $(\pm q_0,0)$. Thus the particles
created by the fundamental fields in the original, \lq\lq electric"
action lie on the real, electric axis. The magnetic monopole solitons
$M^{\pm}$ lie on the imaginary, magnetic axis at $(0,\pm g_0)$, while
the dyons could lie on the horizontal lines through these two points.
Since, at this stage, it is unclear what values of their electric
charges are allowed, they will temporarily be omitted, to be restored later.

Now, if we follow the transformation (14) by a rotation through a
right angle in the $q+ig$ plane, the five points just described  are
rearranged.
This suggests that the \lq\lq dual" or magnetic formulation of the
theory with $M^{\pm}$ created by fields present in the action will
also be a similar spontaneously broken gauge theory, but with the
coupling constant altered by (14). In this new formulation it is the
$W^{\pm}$ particles that would occur as solitons.

This is the Montonen-Olive electromagnetic duality conjecture in its
original form [26]. In principle, it could be proven by finding the
analogue of Skyrme's vertex operator construction (5) [11], but, even with
 present knowledge, this seems impossibly difficult. Notice that the
sine-Gordon quantum field theory was described by two quite dissimilar
actions whereas in the four dimensional theories the two hypothetical actions
have
a similar structure but refer to  electric and magnetic formulations.

The magnitudes of physical quantities should agree whichever of
the two actions is chosen as a starting point for their calculation. The
 conjecture
will immediately  pass at least two simple tests of this kind, showing
 that it is not obviously inconsistent. The first test concerns the
 mass formula (13) and is passed precisely
because of the universal property that has already been emphasised.

A second test concerns the fact that, according to the existence of
static solutions to the Bogomolny equations with magnetic charge
$2g_0$ discussed earlier, an $M^+M^+$ pair exert no static forces on each
other.
This
result is according to the electric formulation of the theory and ought
to be confirmed in the magnetic formulation. This is equivalent to
checking that there is no $W^+W^+$ force in the electric formulation.
In the Born approximation, two Feynman diagrams contribute, photon and
Higgs exchange. Using Feynman rules, one finds that photon exchange
yields the expected Coulomb repulsion but that the second contribution
precisely cancels the first. This can happen because the Higgs is massless
in the Prasad-Sommerfield limit [26].

Thus, at the level considered, the conjecture is consistent, but there
are more searching questions to be asked. Their answers will lead to a
reformulation of the conjecture that passes even more stringent tests.
\bigskip
\ni{\bf Catechism concerning the Duality Conjecture}
\bigskip
The Montonen Olive electromagnetic duality conjecture immediately provokes the
following
 questions:-

\ni(1) How can the magnetic monopole solitons possess the unit spin necessary
for heavy gauge particles?

\ni(2) Will not the quantum corrections to the universal mass formula
(13) vitiate it?

\ni(3) Surely the dyon states, properly included, will spoil the
picture just described?

Clearly the answers to the first two questions will depend on the
choice of quantisation procedure, and presumably the most favourable
one should be selected. The idea of what the appropriate choice was,
and how it answered
the first two questions came almost immediately, though understanding
has continued to improve until the present. The answer to the third
question remained a mystery until it was decisively answered by Sen
in 1994, as we shall describe [28].

 The immediate response of D'Adda, Di Vecchia and Horsley was the proposal that
the
quantisation procedure be supersymmetric [29]. The point is that the theory
we have described is begging to be made supersymmetric since this can
be achieved without spoiling any of the features we have described.
For example, since the scalar and gauge fields lie in the same,
adjoint representation of the gauge group they can lie in the same
supermultiplet. The vanishing of the Higgs self-interaction implied by
the Prasad-Sommerfield limit is then a consequence of supersymmetry.
Because the helicity change between scalar and vector is one unit, the
supersymmetry is
presumably of the \lq\lq extended" kind, with either $N=2$ or $N=4$
possible. Osborn was the first to advocate the second possibility [30].

The reason supersymmetry helps answer the first question is that, given that
it holds in the full quantum theory, it must be represented on any set of
single particle states carrying the same specific values of the  charges and of
 energy and momentum. This is so, regardless of the nature of the particles,
 whether they are created by quantum fields or arise as soliton states,
 that is, whether or not they carry magnetic charge. When the extended
 supersymmetry algebra with $N$ supercharges acts on massive states,
the algebra is isomorphic
to a Clifford algebra in a Euclidean space with $4N$ dimensions [31].
 This algebra has a unique irreducible representation of
$2^{2N}$ dimensions. This representation includes states whose
helicity $h$ varies over a range $\Delta h=N$ with intervals of $1/2$.
The limits of this range may differ but should not exceed $1$
 in magnitude if the states can be created by fields satisfying
renormalisable equations of motion, according to the standard wisdom.

So, of necessity, the
monopoles carry spin, quite likely unit spin. Secondly, quantum
corrections tend to cancel in supersymmetric theories, essentially
 because the
supersymmetric harmonic oscillator has no zero point energy. This
is relevant to the second question because the small fluctuations
about the soliton profile decompose into such oscillators, with the mass
correction equal to the sum of zero point energies.

The structure of the representation theory raises some questions.
According to the renormalisability criterion, the maximum range
of helicity is $\Delta h=1-(-1)=2$, which is just consistent
 with $N=2$ supersymmetry, but apparently forbids $N=4$ supersymmetry.
Another difficulty concerns the understanding of how the Higgs mechanism
providing the mass of the gauge particles, works in the presence
of supersymmetry. The point is that the expression on the right
hand side of the supercharge anticommutator, $\gamma.P$, is a
 singular matrix when $P^2=0$, that is, for massless states.
As a consequence, the supersymmetry algebra acting on  massless
states is isomorphic to an Euclidean algebra in $2N$ dimensions
 and so now possesses a unique irreducible representation of
 $2^N$ dimensions (the square root of the number in the massive case),
 with helicity range $\Delta h=N/2$. This now accommodates both
 $N=2$ and $N=4$ superalgebras but no more. Indeed it is the reason we
 cannot envisage a more extended supersymmetry, such as $N=8$,
 which requires gravitons of spin $2$, whose interactions are not
 renormalisable.

The question arises of how to reconcile the jump in the dimensions
 of the representations with the acquisition of mass for
 a given field content of scalar and gauge fields. The answer,
due to Witten and Olive [32], is that
something special happens precisely when the Higgs field lies in the
adjoint representation, as we have assumed, and so can
lie in the same supermultiplet as the gauge field. Then the
 electric charge, $q$, occurs as a central charge, providing an
 additional term on the right hand side of the supercharge anticommutator,
thereby altering the structure of the algebra. The condition
for a \lq\lq short" representation, that is, of dimension $2^N$,
is now $P^2=a^2q^2$, rather than $P^2=0$. Thus, providing the Higgs formula (9)
holds, mass can be acquired without altering the dimension of
the irreducible representation. Furthermore, magnetic charge
can occur as yet another additional central charge with the condition for a
\lq\lq short"
representation being simply the universal Bogomolny-Higgs formula
 (13). In particular, this means that this formula now
has an exact quantum status as it follows from the supersymmetry algebra
which is presumably an exact, quantum statement (though there may
be subtle renormalisation effects) [32].
\bigskip
\ni{\bf More on Supersymmetry and $N=2$ versus $N=4$}
\bigskip
The possibility that, unlike the unextended supersymmetry algebra, the
extended ones could be modified by the inclusion of central charges
was originally noted by Haag, \L opuszanski and Sohnius [33], while the
physical
identification of these charges was due to Witten and Olive [32]. The
confirmation of the result involved a new matter of principle.
Hitherto supersymmetry algebras had been checked via the algebra of
transformations of the fields entering the action. But since these
will never carry magnetic charge in the electric formulation, this
method will not detect the presence of magnetic charge in the algebra.
Instead, it is necessary to manipulate all the charges explicitly,
treating them as space integrals of local polynomials in the fields
and their derivatives.

The supersymmetry algebras possess an automorphism (possibly outer)
involving chiral transformations of the supercharges
$$Q_{L,R}^{\alpha}\rightarrow e^{\pm i\phi/2}Q_{L,R}^{\alpha}\qquad
\alpha=1,2\dots N\eqno(16)$$
where the suffices refer to the handedness. When the central charges are
present, this automorphism requires them
to simultaneously transform by (6) (at least in the $N=2$ case). Thus
the electromagnetic duality rotation is now seen to relate to a chiral
rotation of the supercharges, sometimes known as $R$-symmetry.

So far, the theory could possess either $N=2$ or $N=4$ supersymmetry and it
is necessary to determine which, if possible. At first sight, $N=2$ is simpler
 as there are precisely two central charges, $q$ and $g$, as we
 have described, whereas
in $N=4$ there are more. However the $N=4$ theory has one very
 attractive feature, namely that there is precisely one irreducible
representation
of the supersymmetry algebra fitting the renormalisability criterion
$|h| \leq 1$, which, moreover,  has to be \lq\lq short", and therefore
satisfy
the universal mass formula (13) [30]. It follows that
 any dyon state must, willynilly, lie in a multiplet isomorphic to the one
 containing the gauge particles. Correspondingly there is only one
 supermultiplet of fields and, as a result,
the supersymmetric action is unique apart from the values of the coupling
constants.

However, there is an even more compelling reason for $N=4$ supersymmetry
 which emerged some years later. In a series of papers it became apparent
that the Callan-Symanzik $\beta$-function vanished identically in the unique
$N=4$
supersymmetric theory [34,35]. This was therefore the first example of a
quantum
field theory in four dimensions
with this property. The vanishing has at least three remarkable consequences
favourable to the ideas considered:

\ni (1) As $\beta$ controls the running of the coupling constant, its vanishing
 means that the gauge coupling constant does not renormalise. Presumably this
applies in
both the electric and magnetic formulations and it means that there is no
question whether the
Dirac quantisation applies to the bare or renormalised coupling constants,
as these are the same (Rossi [36]).

\ni (2) The trace of the energy momentum tensor is usually
proportional to $\beta$ times a local
quantity and so it should vanish in this theory, indicating that the theory is
exactly conformally
invariant (if $a$ vanishes). Thus the $N=4$ supersymmetric gauge theory
is the first known example of a quantum conformal field theory in four
dimensions, to
be compared with the rich spectrum of examples in two dimensions. Furthermore,
the
Higgs mechanism producing a nonzero value for the vacuum expectation value
parameter $a$ presumably provides an integrable deformation
realising Zamolodchikov's principle in four dimensions [8].  Notice that the
naive idea that
conformal field theories should be more numerous in four rather than two
dimensions seems to be false despite the fact that the conformal algebra
has only fifteen rather than an infinite number of dimensions. Besides the
$N=4$
supersymmetric gauge theory, there are now a few other known
 conformal field theories
in four dimensions, all supersymmetric gauge theories.

\ni (3) Finally, just as the trace anomaly vanishes, so does the axial anomaly.
In fact the two properties are related by a supersymmetry transformation.
This means that the chiral symmetry (16) can be extended to the fields of the
theory
and is an exact symmetry for $N=4$. Thus we have answered an
 earlier question and seen that, indeed,
 electromagnetic duality rotations can be extended to include matter,
 albeit in a very special case.

The second point above, concerning the realisation of Zamolodchikov's
principle in four dimensions via a special sort of Higgs mechanism [8],
raises questions about the nature of \lq\lq integrability" in four
dimensions. As far as is known, the $N=4$ supersymmetry algebra is the
largest extension of Poincar\'e symmetry there, but only provides a
finite number of conservation laws, unlike the infinite number
available in two dimensions. On the other hand, there are, apparently,
monopole/dyon solutions with particle-like attributes (certainly if
the duality conjecture is to be believed). But a complete and direct
proof is lacking, even though the results for like monopoles are
encouraging.

For each value of magnetic charge, the  moduli space of solutions
 to the Bogomolny equations (12)
forms a manifold whose points correspond to static configurations of
distinct monopoles with total energy $a|g|$. The problem of describing
their relative motion was answered by Manton [37], at least if it was slow.
His idea follows from the analogy with a  Newtonian point particle confined
to move freely on a Riemannian manifold. It can remain at rest at any
point of the manifold, but, if it moves, it follows a geodesic on the
manifold determined by the Riemannian metric. He realised that the
moduli spaces of the Bogomolny equations must possess such a metric and saw
how to derive it from the action. Actually it has a hyperk\"ahler
structure which makes it very interesting  mathematically. Moreover,
Atiyah and Hitchin calculated the metric explicitly for the moduli
space with double magnetic charge [38]. This is sufficient to determine the
classical scattering of two monopoles at low relative velocity and
yielded surprisingly involved behaviour, including a type of incipient
breathing motion perpendicular to the scattering plane, visible on a video
prepared by IBM.

Despite these beautiful results, there is no idea of how to describe
relative motion of monopole solitons with unlike charge. The duality
conjecture predicts the possibility of pair annihilation, unlike the
sine-Gordon situation. This is why we say the soliton behaviour is
incompletely understood. It is certainly more complicated than in two
dimensions.
\bigskip

\ni{\bf The Schwinger Quantisation Condition and the Charge Lattice}
\bigskip
The remaining difficulty, one that has been repeatedly deferred, concerns
the dyon spectrum. We know that there exist dyon solutions carrying
 magnetic charge, but we do not know what
 values of the electric charge are allowed. The problem is that
the Dirac quantisation condition (7) does not determine this,
 nor does it respect the electromagnetic duality rotation (6) which is
apparently
so fundamental.

It was Schwinger and Zwanziger who independently resolved the problem [39,24].
They saw that  Dirac's assumption that the monopole
carried no electric charge was unjustified, and responsible for the
difficulties.
Instead, they applied Dirac's argument to two dyons, carrying
respective charges $(q_1,g_1)$ and $(q_2,g_2)$, and found
$$q_1g_2-q_2g_1=2\pi n\hbar,\qquad n=0,\pm1,\pm2,\dots.\eqno(17)$$
This is known (somewhat unfairly) as the Schwinger quantisation condition
and it does now
respect the duality rotation symmetry (6) applied simultaneously to
the two dyons. Notice that it is significant that the group $SO(2)$
has two invariant tensors, the Kronecker delta entering the mass formula (13)
and the
antisymmetric tensor entering (17).

As mentioned earlier, the values of $q+ig$ realised by localised states
 composed of particles should lie at the points of a lattice in the
complex plane. The origin of this lattice structure are the conservation laws
 for charge
and the TCP theorem. The set of allowed values must be closed under
both
 addition and reversal of sign
as these operations can be realised physically by combining states and by
TCP conjugation.

Without loss of generality, it can be assumed that there exist a subset
 of states carrying purely electric charge. As long as magnetic charge exists
(17)
implies that there is a minimum positive value, $q_0$, say. Then the allowed
values of pure
electric charge are $nq_0$, $n\in Z\!\!\!Z$, that is, a discrete one
dimensional
lattice.
Now let us examine the most general values of $q+ig$ allowed by the Schwinger
quantisation
condition, (17). By it, the smallest allowed positive magnetic charge, $g_0$
satisfies
$$g_0={2\pi n_0\hbar\over q_0},\eqno(18)$$
where $n_0$ is a positive integer dependent on the detailed theory considered.
Now consider two dyons with magnetic charge $g_0$ and electric
charges $q_1$ and $q_2$ respectively. By (17) and (18)
$$q_1-q_2={2\pi n\hbar\over g_0}={nq_0\over n_0}.$$
However, as there must consequently be a state with pure electric
 charge $q_1-q_2$, $n$ must be a multiple of $n_0$.
Hence for any dyon with magnetic charge $g_0$, its electric charge
$$q=q_0\left(n+{\theta\over 2\pi}\right),$$
where $\theta$ is a new parameter of the theory which is, in a sense,
angular since increasing it by $2\pi$ is equivalent to increasing $n$ by one
unit.
So
$$q+ig=q_0(n+\tau),$$
where
$$\tau={\theta\over2\pi}+{2\pi in_0\hbar\over q_0^2}.\eqno(19)$$
Repeating the argument for more general states with magnetic charge $mg_0$

$$q+ig=q_0(m\tau+n),\qquad m,n\in Z\!\!\!Z.\eqno(20)$$
This is the charge lattice and it finally breaks the continuous
symmetry (2) and (6) in a spontaneous manner [40]. This lattice has periods
$q_0$ and $q_0\tau$ with ratio $\tau$, (19).
Notice that $\tau$ is a complex variable formed of dimensionless parameters
dependent on the detailed theory. Its imaginary part is positive,
 being essentially the inverse of the fine structure constant.

So far, this part of the argument has been very general, but,
 given a specific theory, an important question for
 electromagnetic duality concerns
the identification of the subset of the charge lattice that can
 be realised by single particle states, rather than multiparticle states.

It is easy to show that, if single particle states obey the universal
mass formula (13), and are stable with respect to any two-body decay into
lighter particles permitted by the conservation of electric and
magnetic charge, then they must correspond to points of the charge
lattice which are \lq\lq primitive vectors".

A point $P$ of the charge lattice is a primitive vector if the line
$OP$ contains no other points of the lattice strictly between $O$,
the origin, and $P$. Thus the only primitive vectors on the real axis
are $(\pm q_0,0)$. Equivalently, a primitive vector is a point given
by (20) in which the integers $m$ and $n$ are coprime (in saying this
we must agree that $0$ is divisible by any integer).

The proof of the assertion is simple: it relies on the fact that the
mass of a particle at $P$ is proportional to its Euclidean distance
$OP$ from the origin, by (13). So, by the triangle inequality, any
particle is stable unless its two decay products correspond to points
collinear with itself and the origin. This is impossible, providing the
original particle corresponds to a primitive vector.

There are an infinite number of primitive vectors on the charge
lattice, for example, all the points with $m=\pm1$ or $n=\pm1$. The
corresponding masses can be indefinitely large. If $m=2$, every second
point is a primitive vector. If $m=3$, every third point fails to be a
primitive vector, and so on.

This result tells us what to expect for the spectrum of dyons, namely that
they correspond to the primitive vectors off the real axis. Since the
mass formula used in this argument is characteristic of supersymmetric
gauge theories as discussed above, it ought to be possible to recover
this result from consideration of the Bogomolny moduli spaces
governing the static soliton solutions. This is what Sen achieved in
1994, [28], and a simplified explanation follows.

For $m=\pm1$, the dyons relate to points in the $m=\pm1$ moduli space
since the single monopoles are solutions to the Bogomolny equations.
However, as discussed earlier, the points of the $m=\pm2$ moduli space
correspond to configurations of a pair of like monopoles in static
equilibrium. Thus the $m=\pm2$ single particle states cannot be
Bogomolny solutions. Instead they must be regarded as quantum
mechanical bound states, with zero binding energy (in order to satisfy
the mass formula). Remembering Manton's treatment of moving monopoles
following geodesics on the moduli space determined by the
hyperk\"ahler metric thereon, it is clear that it is crucial to examine the
spectrum of the Laplacian determined by this metric, as this is
proportional to the quantum mechanical Hamiltonian [41]. In particular, zero
modes in the discrete spectrum are sought. There is some subtlety,
treated by Sen, concerning the fact that the quantum mechanics
possesses $N=4$ supersymmetry because the metric is hyperk\"ahler, but
using the Atiyah-Hitchin metric, Sen was able to solve for the zero
modes, and show that only every other permitted value of the electric charge
could occur. Thus the dyons with magnetic charge $2g_0$ do indeed
correspond precisely to the primitive vectors on the charge lattice. For higher
values of $|m|$, the explicit metric is not known, but Hodge's theorem
relates the counting of the zero modes of the Laplacian on the moduli
space to its cohomology, which can be determined without knowledge of the
metric. (This argument is said to be due to Segal, unpublished).

These are the results that finally clear up the dyon problem and leave
the electromagnetic duality conjecture in good shape, though a
reassessment will be in order. Before discussing this, we ask whether
the angle $\theta$, occurring in (19), appears explicitly as a
parameter in the
action of the spontaneously broken gauge theory. Witten found the
answer in 1979 [40]. Because the gauge group is non abelian, $SU(2)$, a
term proportional to the instanton number, $k$ can be added to the
action, so that the Feynman weighting factor becomes:
$$exp\left({i\hbox{Action}\over\hbar}\right)\rightarrow
exp\left({i\hbox{Action}\over
\hbar}+{i\tilde{\theta}\over2\pi}k\right).\eqno(21)$$
As $k$ is proportional to an integral of $F\tilde F$ over space time,
it is a surface term which cannot affect the classical equations of
motion, but it does affect the quantum theory. Note that, like
$\theta$, $\tilde{\theta}$ is an angular variable as the theory is
unaffected if it is increased by $2\pi$. In fact the two angles are
indeed equal as Witten showed by an elementary calculation of the
electric and magnetic charges using Noether's theorem. Thus $\theta$ is
what is known as the instanton or vacuum angle.

The above result has another consequence, yet again singling out the
$N=4$ supersymmetric theory as the only viable one for exact
electromagnetic duality. This is because an application of the chiral rotation
(16) to the fermion fields alters the Lagrangian density by an anomalous term
proportional to the axial anomaly $\beta F\tilde F$. This means that the
instanton angle can be altered by a redefinition of the fermion
field, and so has no physical meaning, unless $\beta$, and hence the
axial anomaly, vanishes. This forces us back to the $N=4$ theory, with
the conclusion that only in this theory does the charge lattice really
make sense. Finally note that in this theory the integer $n_0$
occurring in (19) equals $2$. This is because the $N=4$ theory has
only one supermultiplet which includes the gauge particle and hence
must be an $SU(2)$ triplet. No doublets are allowed in $N=4$, unlike $N=2$.
\bigskip
\ni{\bf Exact Electromagnetic Duality and the Modular Group}
\bigskip
Armed with the new insight that the spectrum of single particle
 states correspond to the primitive vectors of the charge lattice,
augmented by the origin, rather than the five points previously
considered, we can see that the original Montonen-Olive conjecture
was too modest. Instead of possessing two equivalent choices of
 action, the $N=4$ supersymmetric gauge theory apparently possesses an infinite
number
of them, all with an isomorphic structure, but
 with different values of the parameters [28].

Roughly speaking, the reason is that it is the
 charge lattice that describes the physical reality.
Choices of action correspond to choices of basis in the lattice,
 that is a pair of non collinear primitive vectors (or, a pair of periods).
As the charge lattice is two dimensional, these choices are
related by the action of the modular group, an infinite discrete group
containing the previous transformation (14).

Let us choose a primitive vector in the charge lattice, represented by
a complex number, $q_0'$, say. Then we may ascribe short $N=4$ supermultiplets
of quantum fields to each of the three points $\pm q_0'$ and $0$. The
particles corresponding to the origin are massless and neutral whereas
the particles corresponding to $\pm q_0'$ possess complex charge $\pm
q_0'$ and mass $a|q_0'|$. We may form an $N=4$ supersymmetric action
with these fields. It is unique, given the coupling $|q_0'|$, apart
from the vacuum angle whose specification requires a second primitive
vector, $q_0'\tau'$, say, non-collinear with $q_0'$. The remaining
single particle states are expected to arise as monopole solitons or
as quantum bound states of them as discussed above.

Since the two non-collinear primitive vectors $q_0'$ and $q_0'\tau'$
form an alternative basis for the charge lattice, they can be
expressed as integer linear combinations of the original basis, $q_0$
and $q_0\tau$:
$$q_0'\tau'=aq_o\tau+bq_0,\eqno(22a)$$
$$q_0'=cq_0\tau+dq_0,\eqno(22b)$$
where
$$a,b,c,d\in Z\!\!\!Z.\eqno(22c)$$
Equally, $q_0\tau$ and $q_0$ can be expressed as integer linear
combinations of $q_0'\tau'$ and $q_0'$. This requires that the matrix
of coefficients in (22a) and (22b) has determinant equal to $\pm1$,
$$ad-bc=\pm1.\eqno(23)$$
By changing a sign we can take this to be plus one. Then the matrices
$$\pmatrix{a&b\cr c&d\cr}$$
form a group, $SL(2,Z\!\!\!Z)$, whose quotient by its centre is called
the modular group. Equation (22a) divided by (22b) yields
$$\tau'={a\tau+b\over c\tau+d}.$$
These transformations form the modular group and preserve the sign of
the imaginary part of $\tau$. This gives the relation between the
values of the dimensionless parameters in the two choices of action
corresponding to the two choices of basis.
It is customary to think of the modular group as being generated by
elements $T$ and $S$ where

$$T:\tau\rightarrow\tau+1\qquad S:\tau\rightarrow-{1\over\tau}$$
According to (19), $T$ increases the vacuum angle by $2\pi$. This is
obviously a symmetry of (21). If the vacuum angle vanishes, $S$
precisely yields the transformation (14) previously considered.

Proof of the quantum equivalence of all the actions associated with
each choice of basis in the charge lattice would presumably require a
generalised vertex operator transformation relating the corresponding
quantum fields. Since these transformations would represent the
modular group the prospect is challenging.

Meanwhile it has been possible to evaluate the partition function of
the theory on certain space-time manifolds, provided that the theory
is simplified by a \lq\lq twisting procedure" that renders it \lq\lq
topological". Vafa and Witten verified that the results indeed
possessed modular symmetry [42].
\bigskip
\ni{\bf Conclusion}
\bigskip
According to the new results reviewed above, it now appears
increasingly plausible that electromagnetic duality is realised exactly
in the $N=4$ supersymmetric $SU(2)$ gauge theory in which the Higgs
field acquires a non-zero vacuum expectation value. This theory is a
deformation of one of the very few exact conformal field theories in
Minkowski space time. The supporting analysis involves an array of
almost all the previously advanced ideas particular to quantum field
theories in four dimensions, but awaits definitive proof.

Despite its remarkable quantum symmetry this theory is apparently not
physical unless further deformed. Seiberg and Witten have proposed
deformations such that enough structure remains as to offer an
explanation of quark confinement, perhaps the outstanding riddle in
quantum field theory [1,43].

More generally, the potential validity of exact electromagnetic
duality in at least one theory means that quantum field theory in four
dimensions is much richer than the sum of its parts, quantum mechanics
and classical field theory. This is because the new symmetry is
essentially quantum in nature with no classical counterpart.
Moreover it relates strong to weak coupling regimes of the theory.
Consequently,
the new
insight opens a veritable Pandora's box whose contents are now
subject to urgent study.

\bigskip

\def\pl{{\it Phys Lett} } \def\np{{\it Nucl Phys } {\bf  B}}
\def\pr{{\it Phys Rev }}

\ni {\bf REFERENCES}
\medskip
\ni 1. N Seiberg and E Witten, \np {\bf 426} (94) 19-52, {\it Erratum} {\bf
B430} (94) 485-486, \lq\lq
Electromagnetic duality, monopole condensation, and confinement in
$N=2$ supersymmetric Yang-Mills theory".

\ni 2. E Witten, hep-th9411102, \lq\lq Monopoles and four-manifolds"

\ni 3. S Donaldson and P Kronheimer, \lq\lq The geometry of four
manifolds", (Oxford 1990)

\ni 4. L Silberstein, {\it Ann d Physik} {\bf 24} (07) 783-784, \lq\lq
Nachtrag zur Abhandlung \"uber \lq\lq elektromagnetische Grundgleichungen
in bivektorieller Behandlung " "

\ni 5. P Higgs, \pr {\bf 145} (66) 1156-1163, \lq\lq Spontaneous symmetry
breakdown without massless bosons"

\ni 6. F Englert and R Brout, {\it Phys Rev Lett} {\bf 13} (64) 321-323, \lq\lq
Broken symmetry and the mass of gauge vector bosons"

\ni 7. TWB Kibble, \pr {\bf 155} (67) 1554-1561, \lq\lq Symmetry breaking in
non-abelian gauge theories"

\ni 8. AB Zamolodchikov, {\it Advanced Studies in Pure Mathematics} {\bf
19} (89) 642-674, \lq\lq Integrable Field Theory from Conformal field
Theory"

\ni 9. RJ Baxter, \lq\lq Exactly solved models in statistical mechanics",
(Academic Press 1982)

\ni 10. JK Perring and THR Skyrme, {\it Nucl Phys} {\bf 31} (62) 550-555,
\lq\lq A model unified field equation"

\ni 11. THR Skyrme, {\it Proc Roy Soc} {\bf A262} (61) 237-245, \lq\lq
Particle states of a quantized meson field"

\ni 12. S Coleman, \pr {\bf D11} (75) 2088-2097, \lq\lq Quantum
sine-Gordon equation as the massive Thirring model"

\ni 13. S Mandelstam, \pr {\bf D11} (75) 3026-3030, \lq\lq Soliton
operators for the quantized sine-Gordon equation"

\ni 14. HW Braden, EF Corrigan, PE Dorey and R Sasaki, {\it Nucl Phys}
{\bf B338} (90) 689-746, \lq\lq Affine Toda field theory and exact
$S$-matrices"

\ni 15. DI Olive, N Turok and JWR Underwood, \np {\bf 401} (93) 663-697,
\lq\lq Solitons and the energy-momentum tensor for affine Toda theory"

\ni 16. PAM Dirac, {\it Proc Roy Soc} {\bf A33} (31) 60-72, \lq\lq
Quantised singularities in the electromagnetic field"

\ni 17. G 't Hooft, \np {\bf 79} (74) 276-284, \lq\lq Magnetic monopoles in
unified gauge theories"

\ni 18. AM Polyakov, {\it JETP Lett} {\bf 20} (74) 194-195, \lq\lq
Particle spectrum in quantum field theory"

\ni 19. P Goddard and DI Olive, {\it Rep on Prog in Phys} {\bf 41} (78)
1357-1437, \lq\lq Magnetic monopoles in gauge field theories"

\ni 20. EB Bogomolny, {\it Sov J Nucl Phys} {\bf 24} (76) 449-454, \lq\lq
The stability of classical solutions"

\ni 21. MK Prasad and CM Sommerfield, {\it Phys Rev Lett} {\bf 35} (75)
760-762, \lq\lq Exact classical solution for the 't Hooft monopole and
the Julia-Zee dyon"

\ni 22. N Manton, \np {\bf 126} (77) 525-541, \lq\lq The force between 't
Hooft-Polyakov monopoles"

\ni 23. B Julia and A Zee, \pr {\bf D11} (75) 2227-2232, \lq\lq Poles with both
magnetic and electric charges in non-Abelian gauge theory"

\ni 24. J Schwinger, {\it Science} {\bf 165} (69) 757-761, \lq\lq A magnetic
model of matter"

\ni 25. S Coleman, S Parke, A Neveu, and CM Sommerfield, \pr {\bf D15}
(77) 544-545, \lq\lq Can one dent a dyon?"

\ni 26. C Montonen and D Olive, \pl {\bf 72B} (77) 117-120, \lq\lq Magnetic
monopoles as gauge particles?"

\ni 27. P Goddard, J Nuyts and D Olive, \np {\bf 125} (77) 1-28, \lq\lq
Gauge theories and magnetic charge"

\ni 28. A Sen, \pl {\bf 329B} (94) 217-221, \lq\lq Dyon-monopole bound states,
self-dual harmonic forms on the multi-monopole moduli space, and
$SL(2,Z\!\!\!Z)$ invariance in string theory"

\ni 29. A D'Adda, R Horsley and P Di Vecchia, \pl {\bf 76B} (78) 298-302
\lq\lq Supersymmetric monopoles and dyons"

\ni 30. H Osborn, \pl {\bf 83B} (79) 321-326, \lq\lq Topological charges
for $N=4$ supersymmetric gauge theories and monopoles of spin $1$"

\ni 31. W Nahm, \np{\bf 135} (78) 149-166, \lq\lq Supersymmetries and
their representations"

\ni 32. E Witten and D Olive, \pl {\bf 78B} (78) 97-101, \lq\lq Supersymmetry
algebras that include topological charges"

\ni 33. R Haag, JT \L opuszanski and M Sohnius, \np{\bf 88} (75) 257-274,
\lq\lq All possible generators of supersymmetry of the $S$-matrix"

\ni 34. S Mandelstam, \np {\bf 213} (83) 149-168, \lq\lq Light-cone
superspace and the ultraviolet finiteness of the $N=4$ model"

\ni 35. L Brink, O Lindgren and BEW Nilsson, \pl {\bf 123B} (83) 323-328
\lq\lq The ultra-violet finiteness of the $N=4$ Yang-Mills theory"

\ni 36. P Rossi, \pl {\bf 99B} (81) 229-231, \lq\lq $N=4$ supersymmetric
monopoles and the vanishing of the $\beta$ function"

\ni 37. N Manton, \pl {\bf 110B} (82) 54-56, \lq\lq A remark on the
scattering of BPS monopoles"

\ni 38. MF Atiyah and NJ Hitchin \pl {\bf 107A} (85) 21-25, \lq\lq Low
energy scattering of non-abelian monopoles"

\ni 39. D Zwanziger, \pr {\bf 176} (68) 1489-1495, \lq\lq Quantum field
theory of particles with both electric and magnetic charges"

\ni 40. E Witten \pl {\bf 86B} (79) 283-287, \lq\lq Dyons of charge
$e\theta/2\pi$"

\ni 41.  G Gibbons and N Manton, \np {\bf 274} (86) 183-224, \lq\lq
Classical and quantum dynamics of monopoles"

\ni 42. C Vafa and E Witten, \np {\bf 431} (94) 3-77, \lq\lq A strong
coupling test of $S$-duality"

\ni 43. N Seiberg and E Witten, \np {\bf 431} (94) 484-550, \lq\lq
Monopoles, duality and chiral symmetry breaking in $N=2$
supersymmetric QCD"

\bye